\documentclass[11pt]{article}
\usepackage{graphicx} 
\usepackage{amssymb} 
\usepackage{color}
\usepackage{amsthm}
\usepackage{tabularx}
\usepackage{float}
\usepackage{tikz}
\usepackage{hyperref}

\usepackage{setspace}
\usepackage{lipsum} 

\usepackage{amsmath}
\usepackage[
    a4paper,
    left=2.5cm,
    right=2.5cm,
    top=2cm,
    bottom=2cm,
    marginparwidth=0cm,
    marginparsep=0cm,
    footskip=1cm
]{geometry}

\title{Bootstrapping the Finiteness of Leigh-Strassler Deformations and Uncovering Hidden Symmetries}
\author{Lucas S. Sousa\textsuperscript{1} \footnote{mail: santos.sousa@unesp.br}}
\begin{document}
\maketitle
\begin{center}
\textsuperscript{1}
Instituto de Fisica Teorica, UNESP-Universidade Estadual Paulista, R. Dr. Bento T. Ferraz 271, Bl. II, Sao Paulo, 01140-070, SP, Brazil
\end{center}
\begin{abstract}
In this paper, we follow a Bootstrap-like approach to determine the most restricted form the finiteness constraint $\mathcal{F}(q,g,h,\kappa)$, which relates the four parameters of $\mathcal{N}=1$ Leigh-Strassler (LS) deformed models, by imposing mathematical and physical conditions. Focusing first on real parameters, we apply these conditions, together with a new symmetry of the superpotential we named ``q-symmetry'', to strongly constrains $\mathcal{F}$. Imposing only these mathematical conditions is enough, for example, to reproduces the \textit{structure} of the one-loop correction and the \textit{exact result} in the planar limit, which are known from the literature. Extending the analysis to complex parameters, we develop a similar method to obtain the more restricted form of $\mathcal{F}$, though the complex case obscures expansions in ``q-invariant'' variables. We also show how an additional pair $(q,h)$ of integrable deformations arises via q-invariance, and verify that the transformed R-matrix satisfies the Yang-Baxter equation. Moreover, we make two ansatz for the coefficients left in free in the finiteness $\mathcal{F}$ for the real parameters, and while it has some defects, it reveals interesting results when compared with literature: the first predicts the pair of integrable deformations derived in \cite{Mansson2010}, while the second ansatz gives the first correction only at fourth loop order $\kappa^8$ \cite{Mansson2010}, which is known to be true in the planar limit. Furthermore, we study the impact of this symmetry on the algebra of the deformed XXZ spin chain and the moduli-vacuum of LS, and find a gauge/gravity interpretation when $h=0$ for the q-symmetry, obtaining the simplest relation between $k$ (from TsT) and $\beta$ ($q = \exp(\pi i \beta)$) to be linear, in agreement with known results for the Lunin-Maldacena-Frolov deformation \cite{Frolov_2005,Lunin_2005}.
\end{abstract}
\newpage
\tableofcontents
\section{Introduction}
The $\mathcal{N}=4$ SYM theory has been the subject of extensive study over the years, attaining a prominent position among field theories. Its remarkable properties, such as integrability, conformal symmetry \cite{Sohnius:1981sn}, the non-perturbative $\mathrm{SL}(2,\mathbb{Z})$ Montonen-Olive duality, and maximal supersymmetry, make it a cornerstone for exploring phenomena in less constrained theories, such as the gauge-theory of QCD in the Standard-Model, though we should emphasize some progress has been made for the last by studying the symmetries of the amplitudes, \cite{Braun1998}, \cite{Lipatov1993}.

The integrability that the theory exhibits has been thoroughly investigated over the past two decades \cite{Beisert_2011}, and it is widely believed that $\mathcal{N}=4$ SYM is exactly integrable, at least in the planar limit (for deviations from the planar limit, see \cite{Kristjansen_2011}). This is further supported by the AdS/CFT conjecture \cite{Maldacena1997, Witten1998}, which proposes that $\mathcal{N}=4$ SYM is equivalent to type IIB string theory with $Q_{F_5} \sim N$ in AdS$_5 \times$ S$^5$. Since it has been shown that the classical string sigma model on the homogeneous spacetime AdS$_5 \times$ S$^5$ is integrable \cite{Bena2003, Kristjansen_2011}, it is also expected that $\mathcal{N}=4$ SYM is integrable in the corresponding strong coupling and planar limit, $\lambda = N g_{\mathrm{YM}}^2 \gg 1$, $N \rightarrow \infty$. Other examples supporting the success of integrability in $\mathcal{N}=4$ SYM include scattering amplitudes computed using integrable methods, which correctly predict the four-loop deviations observed in \cite{Beisert_2007, Bern_2007}, and the BMN limit, which accurately reproduces the matching of anomalous dimensions of operators with the string spectrum via spin-chain methods \cite{Berenstein_2002}. More generally, there appears to be an intrinsic connection between integrable field theories and spin-chain models, which are often integrable via the Bethe Ansatz \cite{Faddeev1996, Minahan2002, Beisert2003}, including the Hubbard model, XXX Heisenberg, and XXZ chains.

These highly successful results in planar $\mathcal{N}=4$ SYM naturally raise the question: could other theories exhibit similarly well-behaved properties? This question has led to a whole subfield devoted to studying deformations of $\mathcal{N}=4$ SYM, aiming for models that preserve some of the original theory's remarkable features. Ideally, such deformations should at least maintain the integrability of the model. In this direction, numerous deformations have been investigated \cite{Lunin_2005, Frolov_2005, Delduc_2014, sfetsos_2015, Smirnov_2017}.

A large class of deformations of $\mathcal{N}=4$ SYM was discovered by Leigh and Strassler \cite{Leigh_1995}. These deformations preserve part of the supersymmetry, resulting in a wide set of $\mathcal{N}=1$ theories, while also maintaining conformal symmetry, since they correspond to marginal deformations. This class of deformations can be characterized by four parameters $(g, q, h, \kappa)$, corresponding respectively to the undeformed coupling, the parameter deforming the algebra, the coupling associated with a new term in the superpotential, and the deformed coupling. The deformation acts on the superpotential, and it was also shown in \cite{Leigh_1995} that two special symmetries are preserved, namely the cyclic property among the scalar fields ($X_i \rightarrow X_{i+1}$) and invariance under $X_i \rightarrow X_i \omega^{i+1}$. These symmetries, together with conformal exactness, imply a finiteness constraint $\mathcal{F}(g,q,h,\kappa)$ relating the four parameters. Among this family of models, some, such as the $\beta$ deformation \cite{Lunin_2005}, are special points, since $\mathcal{F}$ does not receive quantum corrections at any loop order in the planar limit \cite{Mauri_2005}, and they admit a clear interpretation in terms of transformations of the supergravity background \cite{Frolov_2005}. Other deformations have also been shown to be exactly conformal and related to twists \cite{Mansson2010}, although these cases are less well understood from the perspective of gauge gravity duality.

Rossi \cite{Rossi_2005, Rossi_2006} and Mauri \cite{Mauri_2005} showed the exactness of the finiteness constraint in the planar limit, $N \rightarrow \infty$, for the real $\beta$ deformation, suggesting that integrability is preserved by this deformation, at least in the planar limit. Later, Mansson \cite{Mansson2010, Mansson2007} and Bundzik \cite{Bundzik_2006} constructed several different integrable deformations\footnote{As always, in the planar limit.}, for which the finiteness constraint does not receive quantum corrections and which possess an R matrix satisfying the Yang Baxter equation. These deformations are controlled by a pair of parameters $(q,h)$, which are related by
\begin{equation}
\label{intdef}
    \begin{aligned}
        (q,h) = (0,\frac{1}{\bar{h}} ), \quad ((1+p)e^{in}, p e^{im}), \quad (e^{in}, e^{im}).
    \end{aligned}
\end{equation}
We should mention that in \cite{Mauri_2005, Bork_2008, Kazakov_2007}, the authors identified a symmetry involving the $q$ parameter that deforms the algebra in the amplitudes and in the expansion of the finiteness constraint $\mathcal{F}$. This structure will be shown to be crucial in the present case, although their main results were not explicitly based on this symmetry. Moreover, \cite{Mauri_2005}, for example, focused on the $\beta$ deformation, while here we generalize the analysis to any Leigh Strassler type deformation. In addition, the authors of \cite{Bork_2008} expanded the constraint function $\mathcal{F}$ in terms of what we will refer to as old variables, $(g,\kappa,h,q)$, which we argue correspond to an over counting of independent coefficients, a point that will be demonstrated by the constraints imposed by this symmetry. Here, this symmetry is extended and generalized, and we refer to it as the ``q-symmetry'', which will guide the structure of the calculations. In particular, we generalize it to the remaining parameters of the model, without focusing on Feynman diagram computations, although we will later make use of results obtained in the literature via such methods. For instance, while \cite{Kazakov_2007} performed an expansion in the coupling $g$, here we focus primarily on this symmetry, and the expansion series for $\mathcal{F}$ are constructed from objects that are invariant under the q-symmetry\footnote{For real parameters at first, later generalizing the analysis.}, showing how this strongly constrains the form of the finiteness condition.

This paper is organized as follows. In Section~\ref{sec2}, we summarize the main results of the paper. In Section~\ref{math}, we focus on the case of real parameters to introduce this new symmetry and prove that it holds to all orders, from which we progressively derive a set of restrictions on the structure function $\mathcal{F}$. The next Section demonstrates a precise fitting between the constrained structure and known results from the literature, emphasizing how strongly the q invariance constrains $\mathcal{F}$. In Section~\ref{sec3}, we generalize the analysis to complex parameters, highlighting the obstacles introduced by complexification and proposing a method to overcome them in order to obtain the constrained finiteness function. We also study the integrability properties associated with this symmetry. Finally, in Sections~\ref{sec4a} and~\ref{gg d}, we show how this symmetry extends to objects beyond the superpotential $\mathcal{W}$ and the function $\mathcal{F}$, illustrating how it manifests throughout the model. In particular, in Section~\ref{gg d} we further connect the symmetry to the AdS CFT correspondence, establishing an intrinsic relation between the deformation parameter $k$ from TsT transformations and $q$, a result previously observed in the literature \cite{Frolov_2005}, \cite{Lunin_2005}.

\section{Leigh-Strassler deformed model and discussion of the results}
\label{sec2}
We are looking for a constraint relation between the four global parameters $q$, $\kappa$, $h$, and $g$ of the deformed $\mathcal{N}=4$ theory introduced in \cite{Leigh_1995}. The first parameter, $q$, deforms the algebra of the fields in $\mathcal{N}=4$ into a $q$ algebra of the form $[X,Y]_q = XY - qYX$. The second parameter, $\kappa$, is the deformed coupling, which reduces to the undeformed coupling $g$ in the original theory, and the third parameter, $h$, multiplies a new term in the superpotential of the deformed theory. In this section we use the $\mathcal{N}=1$ supersymmetry description of $\mathcal{N}=4$, and the deformed superpotential is \cite{Leigh_1995}
\begin{equation} \begin{aligned}
\label{w}
    \mathcal{W} = \kappa \; \mathrm{tr} \; ( X_1 [X_2 ,X_3]_q + \frac{h}{3} ( \sum_i X_i^3 ) ) + \text{h.c}.
\end{aligned} \end{equation}
where $X_i$ are the chiral superfields in the superspace formalism. The constraint we are seeking to describe can be expressed as a function of the four parameters
\begin{equation} \begin{aligned}
    \mathcal{F}(g,q,h,\kappa)  =0,
\end{aligned} \end{equation}
resulting from the $\beta=0$ conditions for all the parameters, together with the symmetries LS-theory presents.

In the \textit{real case}, $\mathcal{F}$ assumes a general homogeneous form
\begin{equation} \begin{aligned}
\label{fsec1}
    \mathcal{F}(\gamma_1, \gamma_2, \gamma_3, \gamma_4)  = \sum_{i,j}^{ \infty} \sum_{p=0}^{j} d_{[ij] \; p (j-p)} \gamma_2^p \gamma_3^{j-p} \gamma_1^i \gamma_4^j, \quad    d_{00} = 0,
\end{aligned} \end{equation}
where $\gamma_i$ are parameters defined in terms of the old ones.~\eqref{fsec1} together with further conditions as explained in next sections, suggest the first correction in the planar limit should be of order $g^8$, in agreement with results reported in the literature \cite{Aharony_2002}. The coefficients $d_{\dots}$ are not fully determined by the constraints imposed on $\mathcal{F}$, which is expected: ultimately, these parameters depend on the numerical values obtained by summing over all possible Feynman diagrams at all loop orders.

Nevertheless, we can obtain non-trivial results solely from the relation between $\mathcal{F}$ and the parameters. In the Section that follows, we derive a general expression that can be directly compared with results available in the literature. We show that to first order in $g^2$,
\begin{equation}
    \begin{aligned}
        \lambda = N\kappa^2 \left [\frac{\eta_1}{N^2} q +\left (\frac{1}{2} - \frac{\eta_1}{2N^2} \right )q^2 + \left (  \frac{1}{2} - \frac{\eta_1}{2 N^2}  + \Theta_4 h^2 \right )\right ] ,
    \end{aligned}
\end{equation}
where $\eta_1$ and $\Theta_4$ are the only free coefficients that must be determined from Feynman loop calculations. This result, obtained solely from intrinsic conditions, can be directly compared with results in the literature \cite{Aharony_2002}, assuming real variables,
\begin{equation}
    \begin{aligned}
        \lambda = N\kappa^2 \left [ \frac{2}{N^2} q + \left ( \frac{1}{2} - \frac{1}{N^2} \right ) q^2 + \left ( \frac{1}{2} - \frac{1}{N^2} \right )+ \left ( \frac{1}{2} - \frac{2}{N^2} \right ) h^2 \right ].
    \end{aligned}
\end{equation}
Following two different heuristic approaches in which we analyze the planar limit of the finiteness constraint, we obtain interesting characteristic features for each of them. In the first approach, where we make an ansatz for the coefficients in the series, we correctly predict the conformal and integrable pairs of~\eqref{intdef}, although some caveats remain. In the second approach, where we proceed in the opposite direction and instead impose the vanishing condition at~\eqref{intdef} together with another ansatz, and we find that the first correction to $\mathcal{F}$ appears at order $\kappa^8$, consistent with the literature \cite{Penati_2005, Rossi_2005, Bundzik_2006}.

For complex parameters, the analysis becomes more involved, but we can still obtain a constrained function $\mathcal{F}$. We find that it cannot be expressed in terms of the so-called good variables $\gamma_i$, as was possible for the real parameters in~\eqref{fsec1}. Instead, we introduce a systematic method for constructing the structure function in the complex case. We also show that this function is significantly more constrained than one might expect, since a general expansion would allow for three additional parameters $(\bar{q}, \bar{\kappa}, \bar{h})$. To test its predictive power, we compare it with a previous result from the literature, demonstrating that it can be correctly fitted.

We also find a relation between the R matrix of the transformed complex pair $(\tilde{q} = q^{-1},\tilde{h} = -\frac{h}{q})$ and that of the original pair $(q,h)$, showing that it is exactly equal to the inverse of the initial R matrix
\begin{equation}
    \begin{aligned}
        R_{\tilde{q},\tilde{h}} = R^{-1}_{q,h},
    \end{aligned}
\end{equation}
and consequently it satisfies the Yang Baxter equation whenever the original pair does, providing a systematic method for generating integrable pairs from the standard one. We further investigate the properties of this transformation, including its implications for deformed algebras, such as the affine Lie algebra associated with the integrable XXZ spin chain, and demonstrate that the transformed R matrix remains integrable.

The last section is devoted to examining this transformation from the perspective of AdS/CFT \cite{Maldacena1997}, identifying the parameter $k$ in TsT transformations, which on the gravity side maps to the parameter $q$, as anticipated in \cite{Lunin_2005, Frolov_2005}. We obtain, as the simplest solution, a linear relation between the logarithm of $q$ and $k$, $k \sim \log q$, where for the special case of real beta, we obtain a linear relation between $k$ and $\beta$. This provides additional evidence, consistent with results in the literature, that our approach is well founded.

\section{Finiteness constraint}

\subsection{Real parameters}
\subsubsection{Mathematical conditions on the structure of \texorpdfstring{$\mathcal{F}$}{F}}
\label{math}
To obtain the finiteness constraint, we follow an approach analogous to that used in bootstrap analyses \cite{Poland2018}, but applied to the finiteness $\mathcal{F}$, which is defined as the condition for the vanishing of the beta function. We assume a set of mathematical axioms and physical principles, and by imposing them we progressively restrict the allowed shape of $\mathcal{F}$.

\begin{figure}[H]
    \hspace{6cm}
    \includegraphics[width=0.4\linewidth]{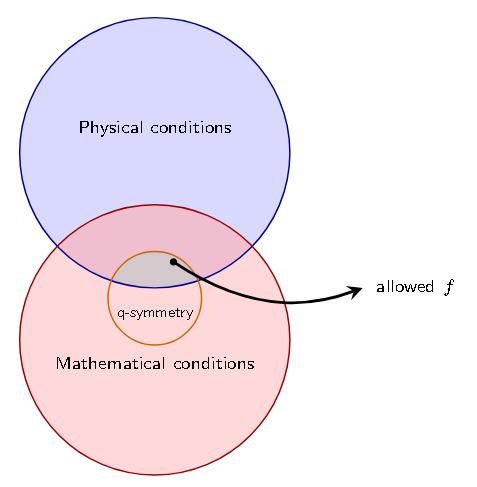}
    \caption{\small{A Venn diagram illustrating the aim of this paper. The regions representing mathematical and physical conditions (e.g., loop effects and consistency) can grow with higher-loop orders in $g$. The allowed region of $\mathcal{F}$ expands with these corrections, but remains far more constrained than generally expected, a ``smaller infinity'' in a sense.
}
}
    \label{i2}
\end{figure}

In principle, $\mathcal{F}$ is intended to describe infinitely many different theories, depending on the relations among the parameters, and our approach highlights this powerful generalization property without specifying any particular quadruple, as illustrated in Figure~\ref{i2}.
First, we demonstrate a discrete symmetry satisfied by the superpotential \\
$\mathcal{W}(g,q,h,\kappa,X_i)$. Consider two models, $1$ and $1'$, both deformed from $\mathcal{N}=4$ SYM and within the LS classification \cite{Leigh_1995}. Each model has a generic deformed superpotential $\mathcal{W}$ and the same matter content. In the unprimed model, we denote the chiral fields as $(X_1, X_2, X_3)$, while in the primed model they are denoted $(X'_1, X'_2, X'_3)$, with the superpotential being
\begin{equation} \begin{aligned}
\label{w1}
    \mathcal{W} = \kappa \; \mathrm{tr} \; ( X_1 [X_2 ,X_3]_q + \frac{h}{3} ( \sum_i X_i^3 ) ),    \quad  \mathcal{W}' = \kappa' \; \mathrm{tr} \; ( X'_1 [X'_3 ,X'_2]_q + \frac{h'}{3} ( \sum_i X_i'^3 ) ).
\end{aligned} \end{equation}
Although not identical in principle, the theories $1$ and $1'$, being subjected to the same type of interactions with identical field content and arranged in exactly the same way, should behave identically under loop corrections. Consequently, the constraint relation among the parameters in $1$, $\mathcal{F}(g,q,h,\kappa)$, and those in $1'$, $\mathcal{F}'(g',q',h',\kappa')$, must be identical,
\begin{equation} \begin{aligned}
\label{rel1}
    \mathcal{F}(g,q,h,\kappa)=\mathcal{F}'(g',q',h',\kappa') =  \mathcal{F}(g',q',h',\kappa').
\end{aligned} \end{equation}
Now, we perform the following relabeling in the $1'$ theory. Since this is a simple relabeling rather than a transformation, all terms in the action remain unchanged,
\begin{equation}
    \begin{aligned}
    \label{transf}
        g' = g, \quad        \kappa' = -\kappa q, \quad        q' = \frac{1}{q}, \quad        h' = -\frac{h}{q} ,     \quad   X'_1 \rightarrow x_1, \quad       X'_2 \rightarrow x_2, \quad        X'_3 \rightarrow x_3,
    \end{aligned}
\end{equation}
then, the superpotential of $1'$ becomes (the other terms of the deformed LS action, which are identical to those of the original $\mathcal{N}=4$ action, remain unaffected by this relabeling),
{\small \begin{equation} \begin{aligned}
    \mathcal{W}' = (-\kappa q)\; \; \mathrm{tr} \; \Big [ x_1 \left(x_3 x_2 - \frac{1}{q} x_2 x_3 \right) - \frac{h}{3q} ( \sum_i x_i^3 ) \Big ] =     \kappa \; \; \mathrm{tr} \; \Big [ x_1 \left(x_2 x_3  - q x_3 x_2 \right) + \frac{h}{3} ( \sum_i x_i^3 ) \Big ] \\
    = \kappa \; \; \mathrm{tr} \; \Big [ x_1 [x_2,x_3]_q + \frac{h}{3} ( \sum_i x_i^3 ) \Big ].
\end{aligned} \end{equation}}
Therefore, the deformed superpotential of $1'$ is identical to that of $1$,~\eqref{w1}, under the simple relabeling $X'_i = X_i$, and hence the Lagrangians of both theories are the same. What is more interesting, and precisely what we are seeking, is the implication of the non trivial transformations~\eqref{transf} for the equality of the constraints~\eqref{rel1}. By applying~\eqref{transf} to~\eqref{rel1}, we find that $\mathcal{F}$ undergoes a non trivial transformation,
\begin{equation} \begin{aligned}
\label{1}
    \bullet\,1.\hspace{4.5cm} 
    \mathcal{F}(g,q,h,\kappa)
    = 
    \mathcal{F}\!\left(g,\tfrac{1}{q},-\tfrac{h}{q},-\kappa q\right)
    \hspace{1 cm}
    \text{``q-symmetry''.}
\end{aligned} \end{equation}
We refer to this as the ``q-symmetry'', since it appears to be induced by the parameter $q$. This provides a strong condition that severely constrains the form of $\mathcal{F}$. Then, $\mathcal{F}$ is constrained to be invariant under such symmetry, and following that line of reasoning we will later seek for new variables that are automatically invariant under all four transformations (one trivial and three non-trivial). For the moment, however, we continue to work with the original variables.

The second and third assumptions state that we can simply change the sign of both couplings, $-\kappa$ and $-g$. This is valid because all terms in the action, aside from those in the superpotential, are quadratic in the fields $X_i$. Therefore, we can perform the transformation $X_i \rightarrow -X_i$ for all three complex fields, while simultaneously taking $\kappa \rightarrow -\kappa$, or $g \rightarrow -g$ in the undeformed case, where the only appearance of $g$ is inside $\tau$, leaving the superpotential invariant. The implication of this transformation for the constraint $\mathcal{F}$ is
\begin{equation}
\begin{aligned}
\label{2}
    \bullet\,2.\hspace{5cm}
    \mathcal{F}(g,q,h,\kappa)
    = 
    \mathcal{F}\!\left(g, q , h , - \kappa\right)
    \hspace{2cm}
    \text{sign 1},\\
     \bullet\,3.\hspace{5cm}
    \mathcal{F}(g,q,h,\kappa)
    = 
    \mathcal{F}\!\left(-g,q,h,\kappa \right)
    \hspace{2cm}
    \text{sign 2}.
\end{aligned}
\end{equation}
Next, we perform a change of variables, $q_i \rightarrow \gamma_i$, which are automatically invariant under q-symmetry, as well as under the sign flip transformations 1 and 2,~\eqref{1} and~\eqref{2}. They are
\begin{equation}
\begin{aligned}
    \gamma_1 = g^2 , \quad \gamma_2 = q + \frac{1}{q} , \quad \gamma_3 = \frac{h^2}{q} , \quad \gamma_4 = \kappa^2 q,
\end{aligned}
\end{equation}
and then, we can write $\mathcal{F}$ as $\mathcal{F} = \mathcal{F}(\gamma_1, \gamma_2, \gamma_3, \gamma_4) = \mathcal{F}(\gamma_i)$.

Further, if $g = 0$, then $\kappa = 0$, and none of the analysis carried out here would make sense. Otherwise, we would be computing a loop correction for a model that has no propagators. This implies
\begin{equation} \begin{aligned}
\label{4}
    \bullet\,4.\hspace{6.3cm}
    \mathcal{F}(0, \gamma_2, \gamma_3, 0)
    = 
    0
    \hspace{3cm}
    \text{free theory.}
\end{aligned} \end{equation}
Looking back at the superpotential~\eqref{w}, we see that nothing prevents us from choosing $q=0$. Therefore, $q=0$ lies within the range of possible values for the deformation parameters and consequently corresponds to a specific constraint $\mathcal{F}(\gamma_i)$ among the constants. In other words, the general form of $\mathcal{F}$ can't contain terms that diverge as $q \rightarrow 0$. We therefore impose analyticity,
\begin{equation} \begin{aligned}
\label{6}
    \bullet\,5.\hspace{5.5cm}
    \mathcal{F}(\gamma_1,\gamma_2,\gamma_3,\gamma_4) \text{ is analytic}
    \hspace{3.5cm}
    \text{A.1}.
\end{aligned} \end{equation}
To see how to avoid such singular terms, we assign an order number $\mathcal{O}$ to each of the $\gamma_i$. This order number is related to the strongest power of $q$ as $q \rightarrow 0$, and we find
\begin{equation} \begin{aligned}
\label{o}
    \mathcal{O}_{\gamma_1} = 0, \quad \mathcal{O}_{\gamma_2} = -1, \quad  \mathcal{O}_{\gamma_3} = -1 ,\quad \mathcal{O}_{\gamma_4} = 1.
\end{aligned} \end{equation}
Since $\mathcal{O}_{\gamma_1^i \gamma_4^j} = j$, we need $\mathcal{O}_{c_{ij}}$ to be at most $-j$ in order to keep a non negative order number. The final expression for the complete structure is therefore obtained without invoking any additional physical input beyond the assumptions stated above,
\begin{equation} \begin{aligned}
\label{c4}
    \mathcal{F}(\gamma_1, \gamma_2, \gamma_3, \gamma_4)  = \sum_{i,j}^{ \infty} \sum_{p=0}^{j} d_{[ij] \; p (j-p)} \gamma_2^p \gamma_3^{j-p} \gamma_1^i \gamma_4^j, \quad d_{00} = 0
\end{aligned} \end{equation}
Although the general expression is best suited for illustrating the overall structure, for practical purposes we will use the form given with the explicit form with $\gamma_{2,3}$ is hidden.

We still have one last conditions $\mathcal{F}$ need to satisfies,
\begin{equation} \label{5}
      \bullet\,6.\hspace{7cm}
    \mathcal{F}(\gamma, 2, 0, \gamma) = 0
    \hspace{3.5cm}
    \text{CC.1}.
\end{equation}
This condition essentially means that $\mathcal{F}$ should reproduce a feature of $\mathcal{N}=4$, where the theory is exactly conformal to all orders. Since the deformed model reduces to the undeformed $\mathcal{N}=4$ case when $q = 1$ and $h = 0$, this implies $g \rightarrow \kappa$.

If we impose all the above conditions at first order in $\kappa^2$, we obtain that
\begin{equation} \begin{aligned}
\label{sl}
    \mathcal{F} = (\alpha_2 + \alpha_3 \gamma_2 + \alpha_4 \gamma_3)\gamma_4 - (\alpha_2 + 2 \alpha_3) \gamma_1 = 0.
\end{aligned} \end{equation}
Or, by rewriting~\eqref{sl} in a clearer way,
\begin{equation}
    \begin{aligned}
    \label{re1}
       \mathcal{F} = \left (-2 \alpha_3 + \alpha_0 + \alpha_3 \gamma_2 + \alpha_4 \gamma_3 \right ) \gamma_4 - \alpha_0 \gamma_1 = 0.
    \end{aligned}
\end{equation}
We can also change the constants and open the new variables $\gamma$ in terms of the older to organize~\eqref{re1} as
\begin{equation}
\label{hof}
    \begin{aligned}
        g^2 = \kappa^2 \left [\Theta_3 q + \left (\frac{1}{2} - \frac{\Theta_3}{2} \right ) q^2 + \left ( \frac{1}{2} - \frac{\Theta_3}{2} + \Theta_4 h^2 \right ) \right ].
    \end{aligned}
\end{equation}
This is the further we can go by imposing only the mathematical conditions we discussed.

However, we can go further by making use of the planar limit of the theory and a back-of-the-envelope calculation. The potential $V$ of the model is given by summing the $F$ terms, ignoring the $D$ terms of the Lagrangian,
\begin{equation}
    \begin{aligned}
        V = \sum_i \mathrm{tr} ( |\partial_i W|^2) = \sum_i \mathrm{tr} ( |[\phi_i,\phi_{i+1}]_q + h \phi_{i+2}^2|^2),
    \end{aligned}
\end{equation}
and we can use only one term of the sum, since it is related to the others by cyclic permutations. Considering $\mathrm{tr} ( |[\phi_1,\phi_2]_q + h \phi_3^2|^2)$, we collect the $q$, $|h|^2$, and $q^2$ terms to obtain
\begin{equation}
\label{lagrang}
    \begin{aligned}
        \mathrm{tr}\!\left(|q|^2 (\phi_2 \phi_1 \phi_1^{\dagger} \phi_2^{\dagger}) - q (\phi_2 \phi_1 \phi_2^{\dagger} \phi_1^{\dagger}) + |h|^2 \phi_3 \phi_3 \phi_3^{\dagger} \phi_3^{\dagger} \right) \subset \mathcal{L},
    \end{aligned}
\end{equation}
where the term involving $\bar{q}$ is analogous to the one involving $q$, and we do not consider it separately. We see that, under Wick's theorem, $|h|^2$ and $|q|^2$ stands out as they contain neighborhood contractions $\phi_1 \phi_1^{\dagger}, \phi_3 \phi_3^{\dagger}$, in opposite to $q$

The handwaving check above has a practical consequence for us. Since the first-loop calculations of $\mathcal{F}$ shouldn't be identically zero, should reduce to $\mathcal{F} = \lambda - N g^2$ when $q=1,h=0$ and $q$ is in a smaller order in comparative with $|h|^2$ and $q|^2$, then necessarily the correct leading power of the terms differs by a factor of $1/N^2$. Then, the finiteness becomes
\begin{equation}
    \begin{aligned}
        \lambda = N\kappa^2 \left [\frac{\eta_1}{N^2} q + \left (\frac{1}{2} - \frac{\eta_1}{2N^2} \right ) q^2 + \left ( \frac{1}{2} - \frac{\eta_1}{2 N^2} \right ) + \Theta_4 h^2 \right ] + O(N^{-3}).
    \end{aligned}
\end{equation}
A second information we can extract from~\eqref{lagrang} is that, at least in the planar limit, the term involving $|h|^2$ and the term involving $|q|^2$ should be related. In fact, this is expected, since they contribute identically to~\eqref{lagrang}, as is clear if we write the $|q|^2$ and $|h|^2$ term as
\begin{equation}
\label{cont}
    \begin{aligned}
        \mathrm{tr} \left ( T_a T_b T_c T_d \right )\left( |q|^2 \phi_2^a \phi_1^b \phi_1^{\dagger c} \phi_2^{\dagger d} + |h|^2 \phi_3^a \phi_3^b \phi_3^{\dagger c} \phi_3^{\dagger d} \right ).
    \end{aligned}
\end{equation}
From~\eqref{cont} we can conclude that, to first order and in the planar limit, $|h|^2$ and $|q|^2$ contributes the same for $\mathcal{F}$, this is because it doesn't matter if $\phi_3$ are the same in the second term, as in the planar limit the only thing that is important is the neighborhood. However, this is not true for the first correction to the planar order: In that case, we allow to $\phi_3^{a}$ pass over $\phi_3^b$, and then we can have twice contractions from the second term, $\delta^{bc} \delta^{ad} + \delta^{ac} \delta^{bd}$, in comparative with the first term, where we only have $\delta^{bc} \delta^{ad}$.

Then, on the finiteness constraint, this implies that
\begin{equation}
\label{our}
    \begin{aligned}
        \lambda = N\kappa^2 \left [\frac{\eta_1}{N^2} q + \left (\frac{1}{2} - \frac{\eta_1}{2N^2} \right ) q^2 + \left ( \frac{1}{2} - \frac{\eta_1}{2 N^2} \right ) + \left ( \frac{1}{2} - \frac{2 \eta_1}{2 N^2} \right ) h^2 \right ] .
    \end{aligned}
\end{equation}
We stress that although the calculations here were carried out for real parameters, we can similarly do it for complex parameters.

For $N$ going to infinite,~\eqref{our} reduces to
\begin{equation}
\label{hof2}
    \begin{aligned}
        \lambda = \frac{N\kappa^2}{2} \left [1 + |q|^2 +|h|^2 \right ] .
    \end{aligned}
\end{equation}
And then, we were able to obtain exactly the planar limit, to first order, of $\mathcal{F}$.

This remarkable result, obtained solely by imposing intrinsic conditions together with some hand waving calculations, can be compared with results from the literature. \cite{Aharony_2002} obtained an expression that gives exactly the same large $N$ limit at first order, but the matching between our result and theirs is not restricted to this large $N$ limit. There, the authors used the superpotential
\begin{equation}
\label{ahar}
    \begin{aligned}
        \mathcal{W} = i \frac{\eta \sqrt{2}}{3!} \epsilon_{ijk} \; \mathrm{tr} \;(\phi^i [ \phi^j,\phi^k]) + \frac{h_{ijk}}{3!} \; \mathrm{tr} \;(\phi^i \{\phi^j, \phi^k \}),
    \end{aligned}
\end{equation}
and by assuming that the only non zero components of $h$ are $h_{111}=h_{222}=h_{333}=h \kappa$ and $h_{123}$, together with the rescaling $\eta = \frac{\kappa(1+q)}{i 2 \sqrt{2}}$ and $h_{123} = \frac{\kappa(1-q)}{2}$, the superpotential reduces to our expression~\eqref{w}. The authors obtained a loop correction of the form $g^2 = \epsilon^2 + \frac{\alpha^2}{2}\left(\frac{-1}{2} + \frac{2}{N^2}\right)$, but we should note that their coupling $g$ does not match our convention. This can be seen, for instance, from their NSVZ correction \cite{Novikov1983}, which agrees with the standard literature only after the rescaling $g = i \frac{g}{\sqrt{2}}$. Substituting this into their correction expression, together with the expressions for $\epsilon$, $h_{123}$, and $h$, their result expressed in terms of our parameters, if we assume $q = \bar{q}$, is
\begin{equation}
    \begin{aligned}
        \lambda = N\kappa^2 \left [ \frac{2}{N^2} q + \left ( \frac{1}{2} - \frac{1}{N^2} \right ) q^2 + \left ( \frac{1}{2} - \frac{1}{N^2} \right ) + \left ( \frac{1}{2} - \frac{2}{N^2} \right ) h^2 \right ],
    \end{aligned}
\end{equation}
to be compared with our result~\eqref{our}. We see that the Feynman diagrams only have the real parameters $\eta_1 = 2$. The fact that our highly constrained function $\mathcal{F}$ is nevertheless able to reproduce the results found in the literature, which in our analogy is similar to fitting a theoretical prediction to experimental data, strongly suggests that our method is correct.

So, returning to the general discussion for $\mathcal{F}$ to any order, we can split the series defining it into three series, those with powers of $\kappa$ higher than two, those with only powers of $g$ higher than two, and the expression obtained previously in~\eqref{sl}, which we denote here by $\mathrm{F}_{g^2,\kappa^2}$,
\begin{equation}
\label{sl2}
    \begin{aligned}
        \mathcal{F}(\gamma_i) = \mathrm{F}_{g^2,\kappa^2} + \sum_i c_{1}^i(\gamma_{2,3}) \gamma_1^i + \sum_{jk} c_2^{jk}(\gamma_{2,3}) \gamma_1^{j} \gamma_4^{k}.
    \end{aligned}
\end{equation}
Imposing CC.1 on~\eqref{sl2}, $\mathcal{F}$ should reduce to the simple constraint of $\mathcal{N}=4$, namely $g^2 = |\kappa|^2$. As we have seen, $\mathrm{F}_{g^2,\kappa^2}$ already reproduces this relation, and the expression above therefore reduces to
\begin{equation}
\label{sl2b}
    \begin{aligned}
        0 = \sum_i c_1^i(\gamma_{2,3}) \gamma_1^i + \sum_{jk} c_2(\gamma_{2,3})^{jk} \gamma_1^{j} \gamma_4^{k}.
    \end{aligned}
\end{equation}
Now, for this equality to be satisfied, we observe the following. While $c_1(\gamma_{2,3})$ is, a priori, a function of $\gamma_2$ and $\gamma_3$, analyticity together with the order counting discussed above implies that the terms in the pure $\gamma_1$ series cannot depend on either $\gamma_2$ or $\gamma_3$. Therefore, to satisfy the equality above, the coefficients $c_2$ need to satisfy  $c_2(\gamma_2 = 2, \gamma_3 = 0)$ and $c_{1}^i = 0$. Therefore,~\eqref{sl2} can be written as
\begin{equation}
\label{sl2c}
    \begin{aligned}
        \mathcal{F}(\gamma_i) = \mathrm{F}_{g^2,\kappa^2} + \sum_{jk} c_2(\gamma_{2,3}) \gamma_1^{j} \gamma_4^{k}.
    \end{aligned}
\end{equation}

\subsubsection*{Higher order terms}

After all the discussion in this paper, one might worry that the q-symmetry is exact only at tree level, that is, at the classical level, and that loop effects start to break it. We have demonstrated that this is not the case, and that the symmetry is genuinely a quantum symmetry, since it acts on complex scalar fields and is a symmetry of the superpotential $\mathcal{W}$, which by itself is non renormalizable. Nevertheless, we can still compare with results from the literature to test whether our expansion of $\mathcal{F}$ can be fitted to known expressions, since this would indicate that our ansatz is correct.

In \cite{Rossi_2005}, the authors calculated a higher order correction to the propagator of a pair of fields. In that analysis, the authors assumed $h=0$, and therefore our expression for the coefficients of $\mathcal{F}$ simplifies drastically to an expansion in $\gamma_2$. The correction they obtained, in our notation and assuming real variables, was
\begin{equation}
\label{paperot}
    \begin{aligned}
        \kappa^2 \Big|_{6} = \kappa^6 g^6 (q - 1)^2 \left ( ( ( q^2 + q + 1 )^2 - 9 q^2 ) N^2 + 5 (q - 1)^4 \right ) \frac{(N^2 - 4)}{N^3}.
    \end{aligned}
\end{equation}
For $\kappa^6 g^6 = \frac{1}{q^3} \gamma_4^3 \gamma_1^3$, the corresponding component in our framework, with $\gamma_3 = 0$, is
\begin{equation}
    \begin{aligned}
        (a_1 + a_2 \gamma_2 + a_3 \gamma_2^2 + a_4 \gamma_2^3 ) \gamma_4^3 \gamma_1^3.
    \end{aligned}
\end{equation}
Imposing CC.1 gives $a_4 = -\frac{a_1 + 2 a_2 + 4 a_3}{8}$. We then find a solution,
\begin{equation}
    \begin{aligned}
    \label{a}
        \alpha_1 \to \frac{8 \left( 20 - 13 N^2 + 2 N^4 \right)}{N^3}, \quad
        \alpha_2 \to -\frac{12 \left( 20 - 9 N^2 + N^4 \right)}{N^3}, \quad
        \alpha_3 \to -\frac{30 \left( -4 + N^2 \right)}{N^3}.
    \end{aligned}
\end{equation}
However, as argued in \cite{Rossi_2005}, the constants can be adjusted so that the $g^6$ correction becomes a $g^8$ correction. We do not pursue this point further here, since the previous analysis was restricted to real parameters. We emphasize a potential source of confusion. The terms obtained in $\mathcal{F}$ are also the most general terms that arise from loop calculations after summing all amplitudes in the same order. In this sense, each term in $\mathcal{F}$ must be consistent with standard Feynman diagram computations. Nevertheless, the expression~\eqref{paperot} is not the final form that should appear in the full sum defining $\mathcal{F}$. This can be seen, for example, by noting that the planar limit of~\eqref{paperot} does not vanish for the $\beta$ deformation.

\subsection{Complex parameters}
\label{sec3}

Since we did not impose any restriction on whether the parameters take values in $\mathbb{R}$ or $\mathbb{C}$ when demonstrating the invariance of the superpotential $\mathcal{W}$~\eqref{w}, the q-symmetry should, and indeed does, hold for both real and complex values. However, in the analysis that followed this demonstration, we restricted the parameters to lie in the real field $\mathbb{R}$. We now justify this choice by showing how the calculations become considerably more involved over $\mathbb{C}$, and afterwards we suggest a less direct approach to the function $\mathcal{F}$.

Following the analysis of Section~\ref{math}, one might naively attempt to construct variables $\gamma_i$ that are invariant under q-symmetry, which is the same symmetry as before but now extended to complex values. Before doing so, however, we observe that there is an additional requirement on the structure of $\mathcal{F}$, namely that it must be real. Suppose we have a finiteness condition of the general form
\begin{equation}\begin{aligned}
\label{g eq}
    \mathcal{F} = \sum_i c_i g^{2i} = 0,
\end{aligned}\end{equation}
with $c_i$ depending on the complex variables $(q,\bar{q},\kappa,\bar{\kappa},h,\bar{h})$. If~\eqref{g eq} holds, then so does its complex conjugate, recalling that $g$ is the only real variable by definition,
\begin{equation}\begin{aligned}
    \label{g eq2}
    \bar{\mathcal{F}} = \sum_i \bar{c_i} g^{2i} = 0.
\end{aligned}\end{equation}
Subtracting~\eqref{g eq} and~\eqref{g eq2}, we find that either $\bar{c}_i - c_i = 0$ for all $i$, or $g = 0$. We of course select the first option, and therefore the coefficients $c_i$ are real, which implies that $\mathcal{F}$ is real. We refer to this property as Reality R.1,
\begin{equation} \begin{aligned}
\label{8}
    \bullet\,7.\hspace{6.2cm}
    \mathcal{F}(g,\alpha_i) \text{ is real}
    \hspace{4cm}
    \text{R.1}.
\end{aligned} \end{equation}
We now try to identify good variables in which to expand $\mathcal{F}$, paying particular attention to conditions R.1 and q-symmetry, which are the strongest constraints in this case. The simplest variables one might guess are
\begin{equation}
    \begin{aligned}
    \label{gamma}
    \Gamma_1 = g^2, \quad
    \Gamma_2 = \frac{1}{2} \left ( c_1 (q + q^{-1}) + \bar{c}_1 (\bar{q} + \bar{q}^{-1}) \right ) + c_2, \\
    \Gamma_3 = \frac{h \bar{h}}{2} \left ( c_3 q^{-1} + \bar{c}_3 \bar{q}^{-1} \right ), \quad
    \Gamma_4 = \frac{\kappa \bar{\kappa}}{2} \left ( c_4 q + \bar{c}_4 \bar{q} \right ).
\end{aligned}
\end{equation}
The difficulty with complex variables arises when imposing analyticity A.1. By treating each complex variable as independent, it is not possible to obtain a simple expression that is simultaneously non zero and non singular in $(q,\bar{q})$. For example, one might expect an expression of the following form at lowest order,
\begin{equation}\begin{aligned}
    \mathcal{F} = \gamma_1 \alpha + \Gamma_4 ( a_1 + a_2 \Gamma_2 + a_3 \Gamma_3),
\end{aligned}\end{equation}
which, when expanded in terms of the old variables, becomes
\begin{equation}
\begin{aligned}
    g^2 \alpha + \frac{|\kappa|^2}{2} \Bigg (
    a_1 c_4 q
    + a_2 c_4 \left( \frac{c_2 ( q^2 + 1 ) + \bar{c}_2 ( |q|^2 + q \bar{q}^{-1} )}{2} \right )
    + a_3 c_4 \left ( \frac{|h|^2}{2} \left ( c_3 + \bar{c}_3 q \bar{q}^{-1} \right ) \right ) \Bigg ) \\
    + \frac{|\kappa|^2}{2} \Bigg (
    a_1 \bar{c}_4 \bar{q}
    + a_2 \bar{c}_4 \left( \frac{c_2 ( |q|^2 + \bar{q} q^{-1} ) + \bar{c}_2 ( \bar{q}^2 + 1 )}{2} \right )
    + a_3 \bar{c}_4 \left ( \frac{|h|^2}{2} \left ( c_3 \bar{q} q^{-1} + \bar{c}_3 \right ) \right ) \Bigg ).
\end{aligned}
\end{equation}
To eliminate the singular terms, and since the $c_i$ are constants, we must impose $c_2 = \bar{c}_2 = c_3 = \bar{c}_3 = 0$, which contradicts the implications of the reductionism condition R.2. This singular behavior persists at any loop order. The origin of the problem can be understood by noticing that the variables are, in analogy with complex analysis or group theoretic language, of degree $(i,0) + (0,i)$. Consequently, any attempt to use $\Gamma_4$ to get rid of the singularity-arising due to the symmetries works only for certain terms, but not for all of them. One might try to redefine the variable $\Gamma_4$ to have degree $(i,i)$, for instance as $(q + \frac{1}{q})(\bar{q} + \frac{1}{\bar{q}})$, or to modify the other variables accordingly, however, this approach is incorrect for two reasons. First, it does not reduce to the real variables $\gamma_i$, thus violating the reductionism condition. First, it significantly increases the power of the variables, leading to higher order corrections already at lowest order, which is not expected. Second, and most importantly, it fails to match previous results from the literature, which at the bootstrap level play a role analogous to experimental validation.

Our situation might appear hopeless, since we do not have good variables beyond the original ones with which to perform the expansion properly. Even though this seems to be the case, we show that it is still possible to obtain a constrained finiteness function $\mathcal{F}$ by following a simple prescription.

Before proceeding, we observe a symmetry under which the superpotential $\mathcal{W}$, together with its conjugate $\bar{\mathcal{W}}$, is invariant. This symmetry was not apparent in the real case because it is intrinsically tied to the complex nature of the variables. We refer to it as the Hermitian condition H.1, and now demonstrate its origin. As is well known, the Leigh Strassler action can be written as
\begin{align}
    S_{\mathcal{N}=4} + \underbrace{\int d^2 \theta \, \mathcal{W}(g_i,X_i) + \int d^2 \bar{\theta} \, \bar{\mathcal{W}}(g_i,\bar{X}_i)}_{S_W} = S_{\text{LS}}.
\end{align}
The action is Hermitian, as in almost all physical models, and therefore it is self conjugate, as is the Lagrangian $\mathcal{L} = \mathcal{L}^{\dagger}$. This has direct implications for the path integral, since
$Z[g_i] = \int D \phi D \bar{\phi} \, e^{-S[\phi,\bar{\phi},g]}$
is mapped to
$\overline{Z[g_i]} = \int D \phi D \bar{\phi} \, e^{-S[\phi,\bar{\phi},\bar{g}]} = Z[\bar{g}_i]$.
In terms of observables, this implies $\langle O \rangle_{g} = \overline{\langle O \rangle}_{\bar{g}}$. Since observables, or expectation values, are themselves subject to Reality R.1, this further implies $\langle O \rangle_{g} = \langle O \rangle_{\bar{g}_i}$, and therefore
\begin{equation} \begin{aligned}
\label{9}
    \bullet\,8.\hspace{6.2cm}
    \mathcal{F}(g,\alpha_i) = \mathcal{F}(g,\bar{\alpha}_i)
    \hspace{4cm}
    \text{H.1}.
\end{aligned} \end{equation}

\subsubsection*{Procedure for complex variables}

The method to obtain the finiteness constraint is simple. First, we treat all variables as real variables, so that we can expand $\mathcal{F}$ in terms of the $\gamma_i$ and impose all the real conditions on its structure. We then expand the resulting expression in terms of the original variables $q_i$, still treating them as real. Second, we allow the original variables to become complex in the following way,
\begin{equation}
\begin{aligned}
    \kappa^{2n} \rightarrow (\kappa \bar{\kappa})^n, \quad h^{2n} \rightarrow (h \bar{h})^n.
\end{aligned}
\end{equation}
This replacement is always valid, since in the real expansion only even powers of $h$ and $\kappa$ appear. In fact, this prescription automatically satisfies the Hermitian condition H.1 discussed in the previous section. The most laborious step, though not excessively so, concerns the parameter $q$ and its coefficients. We first decompose $q^j$ into all possible combinations of the form $\sum_{i+k=j} c_{ik} q^i \bar{q}^k$, where each term is, in principle, multiplied by an independent coefficient. We then require analyticity in the complex variables and allow only non singular terms. Recalling that the expression must also be invariant under q-symmetry, the coefficients $c_{i_1 i_2}$ are related, with the precise relations depending on the factor of $(\kappa \bar{\kappa})^n$ multiplying each term. For example,
\begin{equation}\begin{aligned}
\label{case1}
    \kappa^6 a_1 q^5 \rightarrow (\kappa \bar{\kappa})^3 \big ( c_1 q^3 \bar{q}^2 + c_1 q^2 \bar{q}^3 \big ) \subset \mathcal{F}.
\end{aligned}\end{equation}
After imposing this requirement on all terms in the expansion of $\mathcal{F}$, the analysis is complete. As an illustration, we now repeat the same exercise carried out before and verify that the results agree.

This can also be used to check with the calculations in \cite{Rossi_2005}, this time not restricted to real parameters. In fact, the exact result obtained there is
{\footnotesize\begin{equation}\begin{aligned}
\label{comprossi}
    \frac{(-4 + N^2)\,(-1 + q)\,(-1 + \bar{q})}{N^3}
    \Big (
    5\,(-1 + q)^2 (-1 + \bar{q})^2
    + N^2 \left( -9\, q \bar{q} + (1 + q + q^2)(1 + \bar{q} + \bar{q}^2) \right)
    \Big )
    (\kappa \bar{\kappa})^3 g^6.
\end{aligned}\end{equation}}
We now follow the prescription described above. First, we write the expression for $\mathcal{F}$ expected in the real case, which is the same as obtained previously and which we reproduce here for the reader's convenience,
\begin{equation}\begin{aligned}
    (a_1 + a_2 \gamma_2 + a_3 \gamma_2^2 + a_4 \gamma_2^3)\, \gamma_4^3,
\end{aligned}\end{equation}
we then expand the invariant variables in terms of the original ones and impose Reality R.1, analyticity A.1, the Hermitian condition H.1, and q-symmetry, obtaining
\begin{equation}\begin{aligned}
\label{cc1}
\propto\,
a_4
+ \frac{a_3}{2}( q + \bar{q})
+ (c_1 q^2 + c\, q \bar{q} + c_1 \bar{q}^2)
+ (c_2 q^3 + c_3 q^2 \bar{q} + c_3 q \bar{q}^2 + c_2 \bar{q}^3)\\
+ (c_1 q^3 \bar{q} + c\, q^2 \bar{q}^2 + c_1 q \bar{q}^3)
+ \frac{a_3}{2} ( q^3 \bar{q}^2 + q^2 \bar{q}^3)
+ a_4 q^3 \bar{q}^3.
\end{aligned}\end{equation}
We thus see that the number of real parameters increases from $(a_1,a_2,a_3,a_4)$ to \\
$(a_4,a_3,c,c_1,c_2,c_3)$, that is, only two additional constants compared with the real case. It is possible to adjust~\eqref{cc1} to fit~\eqref{comprossi} exactly, with the solution
\begin{equation}\begin{aligned}
    c_1 = - \frac{a_3}{2}, \quad
    c_2 = - a_4, \quad
    c_3 = - c,
\end{aligned}\end{equation}
and one can verify it fits well, as expected.

\section{Integrability}

The Yang-Baxter equation for the quantum R-matrix is ~\eqref{yb eq} ~\cite{Slavnov2018},
\begin{equation}
\label{yb eq}
\begin{aligned}
    R_{12} R_{13} R_{23} = R_{23} R_{13} R_{12}.
\end{aligned}
\end{equation}
We then borrow the results already derived in~\cite{Mansson2010} for the R-matrix of general deformations of LS-type. We rewrite it here for the reader's convenience,
\begin{equation}
\label{eq:Mmatrix}
\resizebox{0.90\textwidth}{!}{%
$
R=
\frac{1}{|q|^2+|h|^2+1}
\begin{pmatrix}
1 + q\bar{q} - h\bar{h} & 0 & 0 & 0 & 0 & -2\bar{h} & 0 & 2h\bar{q} & 0 \\
0 & 2\bar{q} & 0 & 1 - q\bar{q} + h\bar{h} & 0 & 0 & 0 & 0 & 2h\bar{q} \\
0 & 0 & 2q & 0 & -2h & 0 & q\bar{q} + h\bar{h} - 1 & 0 & 0 \\
0 & q\bar{q} + h\bar{h} - 1 & 0 & 2q & 0 & 0 & 0 & 0 & -2h \\
0 & 0 & 2h\bar{q} & 0 & 1 + q\bar{q} - h\bar{h} & 0 & -2\bar{h} & 0 & 0 \\
2h\bar{q} & 0 & 0 & 0 & 0 & 2\bar{q} & 0 & 1 - q\bar{q} + h\bar{h} & 0 \\
0 & 0 & 1 - q\bar{q} + h\bar{h} & 0 & 2h\bar{q} & 0 & 2\bar{q} & 0 & 0 \\
-2h & 0 & 0 & 0 & 0 & q\bar{q} + h\bar{h} - 1 & 0 & 2q & 0 \\
0 & -2\bar{h} & 0 & 2h\bar{q} & 0 & 0 & 0 & 0 & 1 + q\bar{q} - h\bar{h}
\end{pmatrix}
$
}.
\end{equation}
Equation~\eqref{eq:Mmatrix} represents a quantum R-matrix in a general form. We should note that we are slightly abusing the notation, since the matrix above satisfies the Yang-Baxter (YB) equation only for specific pairs $(q,h)$, and it would therefore be more appropriate to refer to it as the R-matrix only for these pairs. In any case, if $(\tilde{q},\tilde{h})$ satisfies the YB equation, then the model is quantum integrable. The deformed R-matrix is, in terms of the old one,
\begin{equation}
\begin{aligned}
    \tilde{R}\!\left(\frac{1}{\tilde{q}},-\frac{\tilde{h}}{\tilde{q}}\right)
    = R(\tilde{q},\tilde{h})^{-1}.
\end{aligned}
\end{equation}
This property, namely that the deformed R-matrix can be expressed in terms of the undeformed R-matrix, also holds in the complex case. It is straightforward to see that $R^{-1}$ also satisfies the YB equation~\eqref{yb eq} by recalling the definition of $R_{ij}$~\cite{dye200},
\begin{equation}
\label{ror}
\begin{aligned}
\begin{matrix}
    R_{12} = R \otimes I, \\
    R_{13} = P_{23} R_{12} P_{23}, \\
    R_{23} = I \otimes R
\end{matrix}
\Rightarrow
\begin{matrix}
    R_{12}^{-1} = R^{-1} \otimes I, \\
    R_{13}^{-1} = P_{23} R_{12}^{-1} P_{23}, \\
    R_{23}^{-1} = I \otimes R^{-1}
\end{matrix}.
\end{aligned}
\end{equation}
Thus, if~\eqref{yb eq} is satisfied by~\eqref{ror}, then, by taking the inverse of both sides and using $P_{ij}^{-1}=P_{ij}$, we also obtain a solution.

In fact, this can be seen explicitly by making use of the relations in~\eqref{intdef} and verifying that they are mapped either to each other or to themselves. We illustrate this for the second pair, which is the most natural one to consider under inversion,
\begin{equation}
    \begin{aligned}
        \big ( (1+\rho) e^{i n}, \rho e^{i m} \big )
        \rightarrow
        \left ( \frac{e^{-i n}}{1+\rho}, -\frac{\rho}{1+\rho} e^{i(m-n)} \right )
        \underbrace{=}_{\substack{m-n = M \\ n = -N \\ \frac{1}{1+\rho} = 1+x}}
        \big ( (1+x)e^{i N}, x e^{i M} \big ).
    \end{aligned}
\end{equation}

\section{First discussion of heuristic proposal for the planar limit of the real \texorpdfstring{$\mathcal{F}$}{F}}

As we saw, the bootstrap+back-of-the-envelope calculations are quite powerful, but still not enough to understand the general behavior $\mathcal{F}$ should have, and one asks if we can't go further. Motivated by those questions we remember that perturbations of the first order contribution to $\mathcal{F}$ have further roots~\eqref{intdef}, but only in the planar limit. Then, in the present Section we follow two heuristic approach for $\mathcal{F}$:
\begin{itemize}
    \item We first use a simple ansatz for the coefficients of the planar-limit of $\mathcal{F}$, and see if it predicts some of the roots.
    \item We use an ansatz again, but now also bootstrap the planar-limit of the corrections to $\mathrm{F}$ by imposing it to vanish on~\eqref{intdef}.
\end{itemize}
To approach that problem, we recall the superpotential $\mathcal{W}$ in the form given in \cite{Aharony_2002},~\eqref{ahar}. We saw  there are three cubic interaction vertices, one antisymmetric, $\epsilon f_{abc}$, and two symmetric ones, $h_{ijk} d_{abc}$ and $\frac{h}{3} d_{abc}$. The terms in the constraint $\mathcal{F}$ that contain $\kappa$ (that is, all terms, since we have seen that $\sum_i g^i$ must vanish in comparative with $\sum_i \kappa^i$) necessarily originate from interactions involving these different vertices.

Then, we stress the total contribution to higher order can be written, strategically motivated, as
\begin{equation}
\label{sum}
    \begin{aligned}
        \mathcal{F} = \underbrace{g^2 - \kappa^2 \left ( \frac{1 + q^2 + h^2}{2} \right )}_{\mathrm{F}_{g^2, N \to \infty}} - \sum_{i=2}^{\infty} \mathcal{A}_{2i} + O(1/N),
    \end{aligned}
\end{equation}
where $\mathcal{A}$ is planar perturbation to $\mathcal{F}$, which we called $\mathrm{F}$ as in the previous Sections. $\mathcal{A}$, therefore, starts at order $\kappa^4$. Moreover, $\mathcal{A}$ will be expanded in \cite{Aharony_2002} variables, $(1+q), (1-q)$ and $h$.

Since the condition $q = 1,h=0$ make the $\mathcal{A}$ vanishes, These requirements imply that, for any even $n$,
\begin{equation}
\label{atf}
    \begin{aligned}
        \mathcal{A}_n = \kappa^n \sum_j \sum_{2k=0}^{n} \sum_{2i=2}^{n - k} h^{k} C^{(n)}_{i,j,k} (1+q)^{j} (1-q)^{i} \delta_{j,n-i-k}.
    \end{aligned}
\end{equation}

\subsection{Order by order}

The coefficients $C_{i,j,k}$ are not necessarily correlated, but we make the following assumptions. As we are going to study the large $N$ limit, we may expect in this regime the Feynman diagram and loop calculations simplify considerably, because not only are the relevant contributions of planar order, but also the contractions between $d$ and $f$ become essentially the same\footnote{The contractions involving an odd number of terms are the same up to a factor of the complex unit $i$. However, as we saw, q-symmetry eliminates such terms, and we are left only with even contractions, which are indeed the same.}, as one can see for instance in \cite{deAzcarraga1997}, where the authors manipulate different types of contractions ($\mathrm{tr}(F)^2$, $\mathrm{tr}(F^2)$, etc.) up to sixth order, or by following the appendix of \cite{Haber2019}. Furthermore, the possible interaction vertices are
\begin{equation}
    \begin{aligned}
        f_{abc} \epsilon_{ijk} \phi^i \phi^j \phi^k, \quad d_{abc} h_{ijk} \phi^i \phi^j \phi^k.
    \end{aligned}
\end{equation}
As we saw, the first factor corresponds to the $(1+q) f_{abc}$ interactions, while the last factor splits into $(1-q) d_{abc}$ and $\frac{h}{3} d_{abc}$, where the factor of $1/3$ is compensated, when counting the Feynman diagrams, by acting on the three terms in $\mathrm{tr}(\phi_i^3)$. We can therefore expect that the coefficients in $\mathcal{A}_n$ coming from $(1-q^2)$ and $(1+q)h$ are, at least to some extent, correlated apart from symmetric factors, since from the point of view of Feynman diagrams the interactions are the same, this was explored and correctly verified in the first order correction to $\mathcal{F}$.

Since the interactions $(1-q)(1+q)$ and $(1+q)h$ are the same, one can propose an ansatz for $c$ to investigate the consequences in $\mathcal{A}$. Of course, the ansatz must satisfy certain conditions, and moreover, we should expect to obtain at least a general formula valid for any order $n$. Otherwise, $\mathcal{A}$ would be out of control and we would have no final result.

\subsubsection*{Naive ansatz}

We can examine how $\mathcal{A}$ behaves after assigning to $C$ different functions of $i,j,k$. Of course, the function we assume for $c$ cannot give a zero value in the sum for any $q,\kappa$. We can simulate generic functions to see how~\eqref{atf} behaves for real parameters. To illustrate, for $n = 6$ we have,
{\small\begin{equation}
    \begin{aligned}
        \mathcal{A}_6 =  \kappa^6 \left ( C^{(6)}_{2,4,0} (1+q)^4 (1-q)^2 + C^{(6)}_{4,2,0} (1-q)^4 (1+q)^2 + C^{(6)}_{6,0,0} (1-q)^6 \right ) \\
        + h^4 \left ( C^{(6)}_{0,2,4} (1+q)^2 + C^{(6)}_{2,0,4} (1-q)^2 \right )
        + h^2 \left ( C^{(6)}_{0,4,2} (1+q)^4 + C^{(6)}_{2,2,2} (1+q)^2 (1-q)^2 + C^{(6)}_{4,0,2} (1-q)^4 \right ).
    \end{aligned}
\end{equation}}
As explained, one part of the present ansatz will be to require $C$ to be symmetric in $h$ and $(1-q)$, so
\begin{equation}
    \begin{aligned}
       C^{(n)}_{k,i,j} = C^{(n)}_{j,i,k}.
    \end{aligned}
\end{equation}
We assume a splitting property, such that $C^{(n)}_{x_1,x_2,x_3} = a(x_2)\, d(x_1,x_3)$, with $b$ symmetric in its indices. The different types of $b$ include exponential functions, power functions, and constants, combined in various ways,
\begin{equation}
\label{c}
    \begin{aligned}
        d(i,n)\quad = \quad \begin{matrix}
            a_0 \left(b_0^i + b_0^{(n-i)} \right) , \quad
            \underbrace{a_0 \left(i^N + (n-i)^N \right)}_{N=3,2,1,0} , \quad
            a_0 \binom{(n+i)/2}{i/2} , \quad
            a_0 i(n-i) , \quad
            a_0 \binom{n/2}{i},
        \end{matrix}
    \end{aligned}
\end{equation}
and for $a(x_2)$, we can assume,
\begin{equation}
    \begin{aligned}
       a(x_2) \quad = \quad \begin{matrix}
            \underbrace{c_0 x_2^N}_{N = 2,1,0} , \quad
            c_0 (x_2!)^{\pm 1} , \quad
            c_0 2^{\pm x_2}
        \end{matrix}.
    \end{aligned}
\end{equation}
It is hard to compare how each of them behaves just by looking at the results themselves, since some of them involve a messy combination of parameters, the simplest case being the constant one. However, we can compute a contour plot for each after assigning numerical values $a_0 = c_0 = 1$ and $b_0 = 2$, and then compare the different plots to observe similarities between them. To investigate this, we simulated values of $\kappa$ varying in a given range, and values of $\kappa$ and $q$ over a range with integer steps, summing $\mathcal{F}$ up to the $20$th order in $\kappa$. We collected the points for which the modulus of $\mathcal{F}$ is less than a specific numerical value, depending on the marks, as we explicitly show in the graphic that follows. In total, we performed $49$ simulations, but there are particular plots that stand out among the others. This happens when we have $c(k)$ of linear order or higher, with $d(i,n)$ being the first binomial case in~\eqref{c}.

\begin{figure}[H]
\centering
\begin{tabular}{cc}
\includegraphics[width=0.45\textwidth]{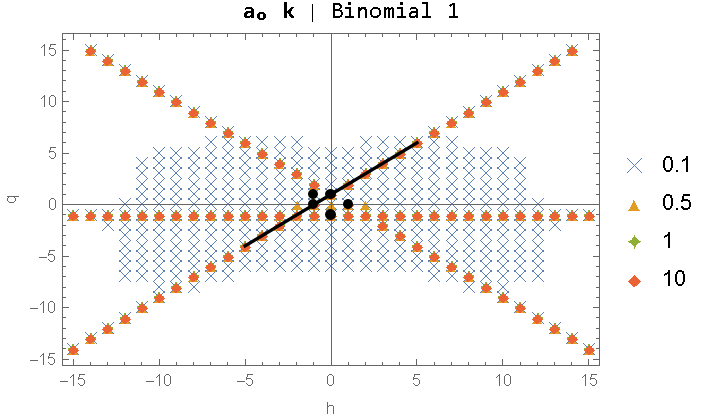} &
\includegraphics[width=0.45\textwidth]{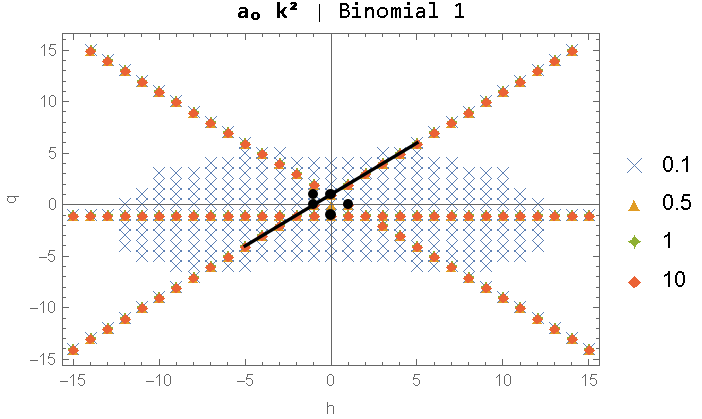}
\end{tabular}
\caption{The black line represents the pairs in~\eqref{intdef}, while the marks in the graph represent the possible roots. The main marks is $\mathrm{X}$, where the function is nearly $0$. } 
\end{figure}

The points and lines in the graphs represent the one-planar finiteness conditions of~\eqref{1}: $(q=-1,h=1)$, $(q=1,h=0)$, $(q=-1,h=0)$, $(q=0,h=1)$, $(q=0,h=-1)$, $q = 1+h$, together with another line $q = 1-h$. The fittings show, to some extent, a prediction for these one-planar finiteness roots.

As we explained, from all the plots we produced, those that stand out are the results from a combination of power functions in $c(k)$ with the first combinatorial case in $d(k)$. We should mention, however, that a similar behavior occurs for any power function of $c(k)$, excluding the constant case (power $=0$), if combined with this specific combinatorial choice. These combinations capture all the discrete roots, along with the relation $q = 1+h$ for any order in $\kappa$, but also suggest that $(q,h) = (-1,h)$ is truly finite for any $h$, a consequence of the choice for $c$. Moreover, the pair $(q=h,1-h)$ was verified to not be integrable, by making use of the same techniques in \cite{Mansson2010} later in the paper, which resembles the limitation of the ansatz. In any case, the others roots were described before, and the coincidence seem to suggest that the asymptotic behavior of the coefficients should be described by the formula $\mathrm{b}_{i,n}\mathrm{c} \sim \exp(k)\binom{\frac{n+i}{2}}{\frac{i}{2}}$, as this includes all powers while discarding the constant term.

\subsubsection*{Bootstrap + ``more refined'' ansatz}

Even though the previous approach was suggestive, in that depending on different combinations of functions we could expect an asymptotic behavior for $c$, the oversimplification was perhaps too trivial. Of course, the amplitudes coming from $h$ and $(1-q)$ are not completely equal. In that direction, one could try to make a better ansatz by noticing that $h$ comes from a vertex with all three fields equal, $\phi_I^3$, and a better fitting happens when, for each loop resulting from $h^{2n}$, we include a $2^{n}$ combinatorial factor compared to $(1-q)^n$, as occurs, for example, in the first-order correction
\begin{equation}
    \begin{aligned}
        \frac{\kappa^2}{2} \left ( c_1 ( 2 h^2 +(1-q)^2) + \frac{1}{2}(1+q)^2 \right ),
    \end{aligned}
\end{equation}
which reduces to the usual expression when $c_1 = 1/2$.

This new approach, together with the ansatz, will be essentially different from the previous one. Instead of guessing a function and selecting those that are more appropriate, we make use of the finiteness pairs in~\eqref{1} to further restrict the coefficients by imposing that each of them should vanish when $(q,h)$ takes any of the values in~\eqref{1}, at each loop order.

We then start with these pair conditions. By imposing the $2^n$ factor explained above together with the splitting property, and also with the imposition of~\eqref{intdef} to be roots of $\mathcal{A}$, and by assuming the same splitting property as before, we obtain an interesting relations between the coefficients $d$, while the $a$ coefficients are independent of each other.

The $d$ coefficients has to satisfy two relations between them
\begin{equation}
\label{d}
    \begin{aligned}
        d_{2n,n} = -\sum_{i=0}^{n-2} d_{2n-2-2i,2n+2+2i} \alpha_i , \quad
        d_{2n,2n+2} = -\sum_{i=0}^{n-2} d_{2n-2-2i,2n+2i+4} \beta_i , \\
        d_{0,k} = 0 \quad d_{2,2} = d_{2,4} = 0,
    \end{aligned}
\end{equation}
where $\alpha$ and $\beta$ are recursive relations,
\begin{equation}
    \begin{aligned}
        \alpha_i = 2^{1+i} + 2^{-(1+i)}, \quad  \beta_i = \frac{1}{6} ( 2^{3+n} + 2^{-n}) .
    \end{aligned}
\end{equation}
These sequences are not new to the literature. Indeed, the first terms of the recurrence are
\begin{equation}
\label{sequences}
    \begin{aligned}
        \alpha = \frac{3}{2} , \frac{11}{4} ,\frac{43}{8} ,\frac{171}{16} ,\dots, \quad \beta = \frac{5}{2}, \frac{17}{4} ,  \frac{65}{8} , \frac{257}{16} , \dots.
    \end{aligned}
\end{equation}
The terms in the numerator of $\alpha$ are tabulated in \cite{OEIS2019}, A007583, and are known to be a type of Lucas sequence, while $\beta$ corresponds to the A092896 sequence. We can make use of a technique known as the binomial transform, which allows us to find an expression for $d$ if we know the inverse sequences of $\alpha$ and $\beta$, which are known to exist. To summarize, suppose we have a sequence $\alpha_i$ that is related to $\beta$ by
\begin{equation}
    \begin{aligned}
        \alpha_j = \sum_{i=0}^{j} \binom{j}{i} \beta_i ,
    \end{aligned}
\end{equation}
then $\alpha$ and $\beta$ are related by a binomial transform, and we can straightforwardly obtain an expression for $\beta$ in terms of the $\alpha_j$,
\begin{equation}
    \begin{aligned}
       \beta_j = \sum_{i=0}^{j} \binom{j}{i} (-1)^{j-i} \alpha_i .
    \end{aligned}
\end{equation}
The fact is that the sequence $\alpha$, A007583, has a binomial transform, which is A025192, while the transform of A092896 is A092897. The sequences in~\eqref{sequences} do not have the initial terms of the cited sequences, and they are also shifted by factors of $2$, but we can straightforwardly fix this by imposing $c_0 = -1$ and by noticing that multiplying all $d_{a,b}$ by $2^{-a}$ (calling this new sequence $u$) does not modify the relations in~\eqref{d} above, since the overall factor of powers of $2$ on both sides will be the same. By doing this,~\eqref{sequences} reduces to the sequences $3,11,43,\dots$ and $5,17,65,\dots$. The fact that we are left without the first terms of the cited sequences is not a problem, and we can find a solution for $u$ in terms of combinatorial factors. First, we define the rule for the sequence transform of $\alpha$, namely A025192,
\begin{equation}
\begin{aligned}
k_{2}(x) = \begin{cases}
1, & x = 0, \\[6pt]
2 \cdot 3^{\,x-1}, & x \neq 0~,
\end{cases}
\end{aligned}
\end{equation}
and for A092897,
\begin{equation}
\begin{aligned}
k_{1}(x) = 
\displaystyle \frac{3^{x} + 4 \cdot0^{x} - (-1)^{x} + 4x(-1)^{x}}{4},
\end{aligned}
\end{equation}
then, $b$ is given completely by four relations,
\begin{equation}
\begin{aligned}
b(M,N) = \begin{cases}
k_{1}\!\left(\frac{M}{2}\right) + (-1)^{\frac{M+2}{2}-1}\!\left(\frac{M}{2}-1\right), & M = N, \\[10pt]
k_{2}\!\left(\frac{M}{2}-1\right) - (-1)^{N/2}, & N = M+2, \\[10pt]
-(-1)^{i}\binom{1+n_{0}}{i}, & 2+2i=M,\ 8+4n_{0}-2i=N, \\[10pt]
-(-1)^{i}\binom{(M+N)/4}{i}, & 2+2i=M,\ 6+4n_{0}-2i=N. \\[10pt]
\end{cases}
\end{aligned}
\end{equation}
For example, $\mathcal{A}$ up to $12$ order is
{\footnotesize\begin{equation}
    \begin{aligned}
        h^2 (1 + h - q) (1 + 2 h - q) (-1 + q)^2 (-1 + h + q) (-1 + 2 h + q) \, \kappa^8 \Bigg(
(1 + q)^2 \kappa^2 \Big( c_2 + \big( c_2 (2 h^2 + (-1 + q)^2) \\
+ c_4 (1 + q)^2 \big) \kappa^2 \Big)
+ c_0 \Big( -1 + (2 h^2 + (-1 + q)^2) \kappa^2 + (4 h^4 + 2 h^2 (-1 + q)^2 + (-1 + q)^4) \kappa^4 \Big)
\Bigg),
    \end{aligned}
\end{equation}}
with the $c_i$ arbitrary.

We see that a first perturbation of \textit{any} pair $(q,h)$ in the \textit{planar limit} is, at minimum, of eight order in $\kappa$. This is expected from the literature, where it is known that the $g^4, g^6$ terms vanish \cite{Penati_2005, Rossi_2005, Bundzik_2006}.

\subsection{General}

In general, only the condition $q = 1 + p$, $h = p$ needs to be imposed at each order. This can be justified by noticing that it is a function of the type $q(h)$, which is more complex than just discrete points like $(-1,1)$ or $(1,0)$. For example, suppose we do not impose vanishing at all orders. Then we have three possibilities. The first is that, at a specific order $n$, the correction is simply zero. The second and third options mimic what happened before: perhaps it is enough to set $q = 1 + p$ or $h = p$ to make it vanish, but this is nonsensical since $p$ is a parameter, and rewriting $q$ and $h$ in terms of $p$ is just a relabeling of parameters. In the end, we would obtain a term that depends neither on $h$ nor on $q$, and therefore is a constant, which necessarily must be zero.

So, either the contribution at that order vanishes, or it vanishes in the specific combination $q = 1 + h$. We adopt the latter approach because, in any case, we can impose the constants, such as $a$ and $c$ before, to vanish. Moreover, if we wanted to impose finiteness up to the third order, we could also solve for the specific cases $\kappa^4(\dots) = \kappa^6(\dots) = 0$. The fact is that, while this can be done, the constraint is much weaker. The first correction, of order $\kappa^8$, already allows for six different solutions, with the coefficients not necessarily correlated among themselves, and therefore the approach tends to be too broad.

We could, nevertheless, make use of a more precise analysis using the Wick contraction, which is the correct approach, to seek a relation between amplitudes involving $(1-q)$ and $h$, but this is not done here and is left for later investigation. Indeed, the previous subsections show that the ansatz method can be iteratively refined, becoming increasingly rigorous, and that, when combined with bootstrap techniques, it could perhaps guide us toward a general form of $\mathcal{F}$, at least in the planar limit.

\section{XXZ spin-chain and holography}

\subsection{Hopf, Yangian Algebra, etc}
\label{sec4a}
The ``$q$-symmetry'' is present at any calculation on the theory of LS-deformed models, and once we identify a correct mapping between this $q$-parameter, the other parameters, and the operators, we can verify that the symmetry is indeed present. The deformation of the algebra involving the fields, $[,] \rightarrow [,]_q$, which now takes values in a deformed gauge group, is well known and has recently been applied in the context of AdS/CFT. The relation between Hopf algebras, or more specifically the Yangian algebra of the deformed group symmetry, and $\mathcal{N}=4$ observables (and consequently AdS$_5\times$S$^5$ as well) has been shown to appear in the S-matrix \cite{Plefka2006}, \cite{Janik2006}, dilatino operators \cite{Gomez2006}, and others. This connection has also been realized and extended to LS-type theories \cite{Mansson2010}, \cite{Dlamini2016}. Here, we follow a less rigorous approach to demonstrate how the ``$q$-symmetry'' is present at the algebraic level, as a preliminary step toward a more complete calculation.

The closed $\mathrm{su}(2)$ sector of the undeformed $\mathcal{N}=4$ theory is a model that, in the planar limit, can be described by a XXX Heisenberg spin chain, with eigenstates built from complex scalars represented by $\uparrow$ and $\downarrow$. In any case, since the algebra is, up to the complex unit $\mathrm{i}$, equivalent to $\mathrm{sl}(2)$, we work with the latter algebra in order to avoid explicitly carrying factors of $\mathrm{i}$. The $\mathrm{sl}(2)$ algebra contains three generators and satisfies the following universal relations,
\begin{equation}
    \begin{aligned}
        \relax [t_1,t_2] = 2 t_2, \;[t_1,t_3] = - 2 t_3, \; [t_2,t_3] = t_1.
    \end{aligned}
\end{equation}
The deformed Hopf algebra introduces three additional operations in the vector space: the coproduct, the counit, and an inverse-like operation. However, we focus only on the first structure together with the R-matrix. In this case, the algebra is generated by the symmetry of the deformed Hamiltonian of the XXX spin chain, or more precisely, of the XXZ spin chain, which differs from the former by having an anisotropic coupling for $\sigma_z$.

The XXZ Hamiltonian and its connection with integrability have been known for decades, see \cite{Jimbo1985}, and more recently \cite{Mansson2007}, \cite{Mansson2010}. The deformed algebra is the quantum group $U_q(\mathfrak{sl}_2)$, and since the $\mathrm{su}$ and $\mathrm{sl}$ algebras are homomorphic, we work instead with $\mathfrak{sl}(2)$ to avoid introducing complex factors. The algebra is \cite{Mansson2010}
\begin{equation}
\label{algsl2}
    \begin{aligned}
        q^\frac{H}{2} X^{\pm} q^{-\frac{H}{2}} = q^{\pm} X^{\pm}, \quad
        [X^+,X^-] = \frac{q^H - q^{-H}}{q - q^{-1}}.
    \end{aligned}
\end{equation}
Under the ``$q$-transformation'', $q \rightarrow \frac{1}{q}$,the above algebra remains the same, and the operators don't transform. The coproduct $\Delta$ of the Hopf algebra is given by
\begin{equation}
\label{hop1}
    \begin{aligned}
        \Delta q^{\pm H/2}
= q^{\pm H/2} \otimes q^{\pm H/2},
\qquad
\Delta X^{\pm}
= X^{\pm} \otimes q^{H/2}
+ q^{-H/2} \otimes X^{\pm}.
    \end{aligned}
\end{equation}
While the first relation remains the same under q-transformation, the coproduct acting on $X^{\pm}$ transforms to
\begin{equation}
\label{hop2}
    \begin{aligned}
        \tilde{\Delta} X^{\pm} = q^{\frac{H}{2}} \otimes X^{\pm} + X^{\pm} \otimes q^{-\frac{H}{2}},
    \end{aligned}
\end{equation}
and one can notice the terms in~\eqref{hop2} are the same as~\eqref{hop1}, swapped however, and consequently, $\tilde{\Delta} = P \Delta$, where $P$ is the permutation operator, but since this is the definition of opposite co-product, $\tilde{\Delta} = \Delta_{\mathrm{op}}$. Then, since $U_q(\mathfrak{{sl}(2))}$ is quasitriangular, we know exists a matrix $R$ connecting $\tilde{\Delta}$ and $\Delta$, with $R$ satisfying the Yang-Baxter equation. This is only one of the many pieces of evidence in the paper where ``q-symmetry'' preserves integrability.

In that XXZ spin-chain, we can also study the ``q-symmetry'' through non-perturbative features, by studying the vacuum conditions of the Hamiltonian. This was done for the $\beta$ deformation in \cite{Lunin_2005}, where the authors analyzed the two possible vacuum branches (conformal and non-conformal, or Coulomb), both from the field-theory and the (super)gravity perspectives. Here, we follow the approach of \cite{Berenstein2000}, where the moduli space $\big (\phi_j \; | \; V(\phi_j) = 0 \big )$ of deformed $\mathcal{N}=4$ models, including the LS deformation, was studied. The vacuum condition is
\begin{equation}
\label{alg}
    [\phi_1, \phi_2]_q = [\phi_2, \phi_3]_q = [\phi_1,\phi_3]_q = 0.
\end{equation}
Moreover, the authors consider $q$ to be an $N$-th root of unity, $q^N = 1$. We show how this algebra ``reacts'' under the ``$q$-symmetry'' by constructing the isomorphic mapping mentioned earlier. Since the algebra is deformed, its representations depend on the space on which it acts. While one could show directly that the representations of the untransformed space $V_q$ are in one-to-one correspondence with those of $V_{\frac{1}{q}}$, we instead follow the results of \cite{Berenstein2000}. The $q$-algebra above admits representations in a vector space where the $\phi_i$ are represented by matrices in a basis of eigenvectors of $\phi_1$ with dimension $N$. We show that there exists a conjugation matrix $U$ that maps the operators $\phi_i$ to their transformed counterparts, while preserving the algebra.

As shown in \cite{Berenstein2000}, the fields $\phi_i$ can be represented as matrices. Assuming a basis of eigenvectors $|a\rangle$ of $\phi_1$, and noting that $(\phi_2)^n |a\rangle$ is an eigenvector of $\phi_1$ with eigenvalue $q^n a$, where $a$ is the eigenvalue associated with $|a\rangle$, one can construct the entire basis of the space and obtain the representations of $\phi_i$ on the vacuum moduli space. These representations are
\begin{equation}
    \begin{aligned}
        \Phi_1 = \begin{pmatrix}
            1 & 0  & 0 & \dots \\
            0 & q & 0 & \dots \\
            0 & 0 & q^2 & \vdots \\
           0 & \vdots  & \ddots & 0 \\
            0 & \dots & 0 & q^{N-1} 
        \end{pmatrix} ,        \Phi_2 = \begin{pmatrix}
            0 & 0 &  \dots & 0 & 1\\
            1 & 0 & \dots & 0 & 0\\
            0 & 1 & \ddots  & 0 & \vdots\\
            0 & 0 & 1 & \ddots & 0\\
            0 & 0 & 0 & 1 &0
        \end{pmatrix},
    \end{aligned}
\end{equation}
with $\Phi_3 = \Phi_2^{-1} \Phi_1^{-1}$, as discussed in \cite{Berenstein2000}. One can easily verify that these matrices satisfy both the conditions above and the algebra~\eqref{alg}. The vector
\begin{equation}
    \begin{aligned}
        |v_i\rangle = \begin{pmatrix} 0 \\ \vdots \\ 1_{\text{i-th line}} \\ \vdots \\ 0 \end{pmatrix}
    \end{aligned}
\end{equation}
has eigenvalue $q^{i-1}$ under $\Phi_1$. Defining $U$ as the matrix that connects $V_q$ and $V_{\frac{1}{q}}$, we have
\begin{equation}
\begin{aligned}
\label{ç2}
    \Phi_1 | v_i \rangle = q^{i-1} | v_i \rangle \Rightarrow  U \Phi_1 U^{-1} U |v_i \rangle = U q^{i-1} | v_i \rangle.
\end{aligned}
\end{equation}
Since we want to implement the transformation $q \rightarrow q^{-1}$ at the level of states, together with the condition $q^N = 1$, we define the matrix $U$ by
\begin{equation}
    \begin{aligned}
    \label{ç1}
        U | v_1 \rangle = |v_1\rangle, \qquad U | v_i \rangle = | v_{N-(i-2)} \rangle,
    \end{aligned}
\end{equation}
and the explicit form of $U$ in this basis is
\begin{equation}
    \begin{aligned}
        U_{ij} =
\begin{cases}
1, & \text{if } i = 1 \text{ and } j = 1, \\[6pt]
1, & \text{if } j > 1 \text{ and } i + j = n + 2, \\[6pt]
0, & \text{otherwise.}
\end{cases}
    \end{aligned}
\end{equation}
Substituting~\eqref{ç1} into~\eqref{ç2}, we find
\begin{equation}
    \begin{aligned}
        U \Phi_1 U^{-1} | v_{N+2-i} \rangle = q^{i-1}| v_{N+2-i} \rangle.
    \end{aligned}
\end{equation}
Defining the transformed matrices as $\tilde{\Phi}_i = U \Phi_i U^{-1}$, we see that in this new basis, for example,
\begin{equation}
    \begin{aligned}
         \tilde{\Phi}_1 = \begin{pmatrix}
            1 & 0  & 0 & \dots \\
            0 & q^{N-1} & 0 & \dots \\
            0 & 0 & q^{N-2} & \vdots \\
           0 & \vdots  & \ddots & 0 \\
            0 & \dots & 0 & q
        \end{pmatrix}.
    \end{aligned}
\end{equation}
As one can verify, $\tilde{\Phi}_1(q) = \Phi_1(q^{-1})$, and the same relation holds for $\tilde{\Phi}_{2,3}$. Therefore, the matrix $U$ maps $V_q$ to $V_{\frac{1}{q}}$, and both describe the same conjugacy class of representations of the algebra. One still needs to verify that $\tilde{\Phi}_i$ satisfy~\eqref{alg}, but this is immediate once one notices that $\tilde{\Phi}_1(q) = \phi_1(q)^{-1}$, $\tilde{\Phi}_2(q) = \phi_2(q)^{-1}$, and $\tilde{\Phi}_3(q) = q^2 \phi_3^{-1}$, further elucidating the structure of the map.

\subsection{Holography interpretation}
\label{gg d}
We can also show how the $q$-transformation manifests itself from the supergravity point of view by obtaining a solution for the requirement of invariance. We decide to follow the most simple case considered here, with $h=0$.

Frolov \cite{Frolov_2005}, Maldacena and Lunin \cite{Lunin_2005}, have shown that deformations such as the $\beta$ deformation, or more general $\gamma$ deformations involving three parameters (three $\beta$ deformations), can be interpreted as TsT transformations of the background. These consist of a T-duality applied to a coordinate, followed by a shift along an isometric coordinate $x_1 \rightarrow x_1 + \gamma x_2$, and then another T-duality along the first coordinate. These transformations, consisting of abelian T-duality and translations, map supergravity solutions to other supergravity solutions, and \cite{Frolov_2005} showed that they are dual to the LS deformation we are studying here (again, with $h=0$ in this specific case). Moreover, TsT transformations have been exhaustively applied to several models, with an astonishing level of success in being dual to general deformations, in some cases integrable \cite{McGough2016}, \cite{giveon2020tbartlst}.

The TsT transformation affects not only the metric of the background, but also the NSNS and RR fields, with general rules already derived in the literature \cite{Buscher1987}, \cite{Imeroni_2008}. However, we are not going to pay too much attention to this part of the calculation (as mentioned above, this is just the beginning of a later calculation and is meant only to convey the general idea).

Type IIA and type IIB supergravity models have an action of the form, ignoring all other fields except the metric and working in the string frame
\begin{equation}
    S = \frac{1}{2 \kappa^2} \int e^{-2\phi} \sqrt{-g} R + \dots,
\end{equation}
but we know from amplitude calculations that, when the dilaton is constant, $g_s = e^{\phi}$, and therefore the action becomes
\begin{equation} \begin{aligned}
\label{action}
    S = \frac{1}{2 \kappa^2} \int \frac{1}{g_s^2} \sqrt{-g} R + \dots.
\end{aligned} \end{equation}
Suppose a basic metric of the form $(-dt^2 + dx^2)$, and apply a TsT transformation to it. Performing a T-duality along $t$ gives
\begin{equation} \begin{aligned}
    ds^2 = -dt^2 + dx^2 + \dots,
\end{aligned} \end{equation}
and the TsT-deformed metric is
\begin{equation} \begin{aligned}
    ds^2 = \frac{-dt^2+dx^2}{1 - k^2} + \dots.
\end{aligned} \end{equation}
Important to our case, however, is the Yang-Mills coupling (obtained from the relation $e^{\phi} = g_s = g_{\mathrm{YM}}^2$, up to overall factors), which transforms as
\begin{equation}
\label{p1}
    \begin{aligned}
        g_{\mathrm{YM}}^4 \underbrace{\to}_{\mathrm{TsT}} \tilde{g}_{\mathrm{YM}}^{4} = \frac{g_{\mathrm{YM}}^4}{1 + (k^2 \mathcal{M})},
    \end{aligned}
\end{equation}
where $\mathcal{M}$ is a determinant of the metric involving the components affected by the TsT transformation, its precise form is not important for the present discussion.

To connect this with the LS parameter, we assume $h=0$ and rewrite the superpotential in a more symmetric form,
\begin{equation}
    \begin{aligned}
        \mathcal{W} = \tilde{\kappa} \mathrm{tr}\left( \phi_1 \left ( \sqrt{q} \phi_2 \phi_3 - \phi_3 \phi_2 \sqrt{q}^{-1} \right ) \right) =\tilde{\kappa} \mathrm{tr}\left( \phi_1 \left ( \tilde{q} \phi_2 \phi_3 - \phi_3 \phi_2 \frac{1}{\tilde{q}} \right ) \right).
    \end{aligned}
\end{equation}
The new parameters $\tilde{q}$ and $\tilde{\kappa}$ are related to $\kappa$ and $q$ by $\tilde{\kappa} = \kappa \sqrt{q}$ and $\tilde{q} = \sqrt{q}$, and they transform under the $q$-symmetry as
\begin{equation}
\label{p2}
    \begin{aligned}
        \tilde{\kappa} \underbrace{\to}_q - \tilde{\kappa}, \quad \tilde{q} \underbrace{\to}_q \frac{1}{\tilde{q}}
    \end{aligned}
\end{equation}
(note that $\tilde{\kappa}^2 = \gamma_2$). In this notation, the $q$-symmetry splits into two independent transformations, with $\tilde{\kappa}$ transforming independently of $\tilde{q}$, and vice versa. This therefore provides a more natural notation for writing the objects of the theory, properly highlighting the $q$-symmetry (in the same sense as writing irreducible representations of a group).

We now have four parameters, $\tilde{\kappa}$, $\tilde{q}$, $k$, and $\tilde{g}_s$, with the first two coming from the field-theory side and the last two from the supergravity side. By the AdS/CFT conjecture, we should expect a relation between them. The first relation we already know from literature
\begin{equation}
    \begin{aligned}
        \tilde{\kappa}^2 = \tilde{g}_{\mathrm{YM}}^2 = 2 \pi \tilde{g}_s,
    \end{aligned}
\end{equation}
We are then left with a relation of the form $k(\tilde{q},\tilde{\kappa})$. There is no precise derivation of this relation, as the duality for the general Leigh-Strassler deformation is not fully understood, but we can make some contact in this direction by using the transformations~\eqref{p1} and~\eqref{p2}. Under~\eqref{p2}, $\tilde{\kappa}^2$ is invariant, and therefore $\tilde{g}_{\mathrm{YM}}^4$ in~\eqref{p1} remains unchanged.

Nevertheless, the parameter $k$ in the context of the TsT transformation is independent of the coupling\footnote{More precisely, $k$ is an independent parameter that affects the coupling. If $k = k(g_s)$, then the transformed coupling $\tilde{\phi} = \phi_0 + \frac{1}{2} \log \left ( \dots \right )$ would become an equation for $\phi(g_s)$, which would contradict this assumption.}, and therefore one may expect $k = k(\tilde{q})$. Following this logic, under the $q$-symmetry~\eqref{p2}, $k$ should transform as $k(\tilde{q}) \to k(\tilde{q}^{-1})$, but constrained in such a way that $\tilde{g}_{\mathrm{YM}}$ in~\eqref{p1} remains invariant. Consequently, we require $k^2$ to be invariant, leading to two possible transformations,
\begin{equation}
\label{p4}
    \begin{aligned}
        k(\tilde{q}) \to k(\tilde{q}^{-1}) =  - k(\tilde{q}), \quad k(\tilde{q}) \to k(\tilde{q}^{-1}) = k(\tilde{q}).
    \end{aligned}
\end{equation}
We expect the function $k(\tilde{q})$ to have only a simple root, since the undeformed case $k=0$ corresponds to $\mathcal{N}=4$, that is, $\tilde{q}=1$, as implied by the above relations. The antisymmetric case under inversion is harder to constrain, while the second case is simpler but it would require a more general function that is not known and would require a series expansion. We can then see the simplest function that satisfies one of the required properties is
\begin{equation}
\label{p3}
    \begin{aligned}
        \mathrm{k}  \propto \ln \tilde{q},
    \end{aligned}
\end{equation}
in which case we choose the first transformation in~\eqref{p4}. Of course, from this solution one could obtain infinitely many functions with higher transcendentality (such as $\ln^3 \tilde{q}$, etc.), but we choose the simplest closed-form solution rather than a series expansion.

We note that, for the real $\beta$ deformation, $\tilde{q} = \exp\left(\frac{i \pi}{2} \beta\right)$, and~\eqref{p3} implies a linear relation between the parameters $\beta$ and $k$,
\begin{equation}
    \begin{aligned}
        \beta \propto  k,
    \end{aligned}
\end{equation}
where we assumed $\beta$ to be in the principal branch of logarithm. Remarkably, this relation is the same as the one expected from the literature \cite{Frolov_2005}, \cite{Lunin_2005}, where the proportionality becomes an equality, $\beta = k$.

We thus see how the $q$-transformation can be understood from the supergravity point of view within gauge/gravity duality. This calculation suggests that, in principle, one should expect a relation between the TsT parameter $k$ and the parameter $q$ of the deformed algebra, in agreement with the literature \cite{Frolov_2005}. The $\beta$ deformation (the dual of TsT) involves a complex parameter, and \cite{Frolov_2005} demonstrated that for $\beta$ complex (instead of real), the dual description in supergravity involves TsT and $\mathrm{SL}(2,\mathbb{R})$, and therefore one should take our calculation with caution, since generalization is not straightforward and then a more careful analysis is required. This is left for future work, and here we merely elucidate how one might guess a relation between the TsT parameter and the $q$ parameter by 
imposing $q$-invariance.

\section{Conclusion}
The fact that the ``q-symmetry'', and others, strongly constrains the shape of $\mathcal{F}$ has shown to provide useful insights, even being enough to obtain exactly the ``tree-level'' on planar limit of the finiteness function. That seems to be a significant step toward obtaining an explicit constraint function $\mathcal{F}(g,q,h,\kappa)$ that correctly predicts the behavior and relations among the constants for any type of Leigh-Strassler deformation. The comparative of our results with those in the literature suggests that our approach is consistent and that $\mathcal{F}$ should be a function of invariant variables $\gamma_i$, at least when working with real parameters, and could be straightforwardly generalized to complex ones. While not all information is extracted from our analysis, we stress how remarkable it is to obtain a general form for the relation among the constants of the theory a priori. Indeed, from the point of view of loop calculations, one would need to perform infinitely many computations at all orders to finally determine the exact form of $\mathcal{F}$, but the conjecture presented here already defines it.

Moreover, that we can obtain another pair of integrable deformations by using this transformation is remarkable, since the seemingly ``non-offensive'' new symmetry of the superpotential turns out to be a generating mechanism for deriving new integrable deformations from those already known.

Regarding the heuristic approaches, it is evident that they have shortcomings, as they contain roots that are not integrable. Integrability is not necessarily a prerequisite for conformality or finiteness~\cite{Mansson2010}, however, it is generally expected that both concepts go hand in hand. For this reason, we are led to believe that the new roots are purely limitations of our simple ansatz. Despite all these counterpoints, it is quite remarkable that we were able to recover some important properties known from the Leigh-Strassler analysis, at least as it is understood today: the $\kappa^8$ term and the integrable pairs~\eqref{intdef}. Therefore, a more rigorous analysis in the future, based on an appropriate ansatz, could perhaps provide at least an approximation to the final form of $\mathcal{F}$.

Finally, the perspective of this $q$-transformation in the context of the gauge/gravity correspondence can be extended by analyzing $h \neq 0$ or performing more rigorous calculations beyond the metric, since the other NS-NS and RR fields are also affected by the TsT transformation, which we leave for future work. As explained, the connection between the TsT parameter $k$ and the LS parameter $q$ is established in the literature for $|q|^2 = 1$, but the more general case for arbitrary $q$ and $h$ is, to the best of our knowledge, not yet well understood. This symmetry could therefore provide new insights into the duality between the general LS deformation and supergravity.

\section*{Acknowledgments}

LS would like to thank FAPESP for the support provided under grant 2023/13676-4. The authors also thank professor H. Nastase and P. Vieira for the useful discussions.

\bibliographystyle{plain} 
\bibliography{TsT-penrose} 
\end{document}